\documentclass{article}
\makeatletter

\usepackage{amstext}
\usepackage{amssymb}
\usepackage{stmaryrd}
\DeclareFontFamily{OT1}{cmtex}{}
\DeclareFontShape{OT1}{cmtex}{m}{n}
  {<5><6><7><8>cmtex8
   <9>cmtex9
   <10><10.95><12><14.4><17.28><20.74><24.88>cmtex10}{}
\DeclareFontShape{OT1}{cmtex}{m}{it}
  {<-> ssub * cmtt/m/it}{}
\newcommand{\texfamily}{\fontfamily{cmtex}\selectfont}
\DeclareFontShape{OT1}{cmtt}{bx}{n}
  {<5><6><7><8>cmtt8
   <9>cmbtt9
   <10><10.95><12><14.4><17.28><20.74><24.88>cmbtt10}{}
\DeclareFontShape{OT1}{cmtex}{bx}{n}
  {<-> ssub * cmtt/bx/n}{}
\makeatother
\usepackage{amsmath,latexsym,amssymb}
\usepackage{haskell}
\usepackage{graphicx}
\usepackage{url}

\newcommand{\omitnow}[1]{}
\newcommand{\co}[2]{#1 \cosep #2}
\newcommand{\cosep}{\mathbin{:\mkern-7mu+}}

\newcommand{\bhc}{\begin{haskell*}}
\newcommand{\ehc}{\end{haskell*}}
\newcommand{\dirac}[1]{\mid\mkern-5mu#1\rangle}
\newcommand{\outpr}{\mathbin{\rangle\mkern-5mu * \mkern-5mu\langle}}
\newcommand{\hsbind}{\gg\!\!=}
\newcommand{\hseq}{\mathbin{=\mkern-5mu=}}
\newcommand{\scapr}{\mathbin{\$*}}
\newcommand{\tenspr}{\mathbin{\langle*\rangle}}
\newcommand{\dotpr}{\mathbin{\langle \cdot \rangle}}

\newtheorem{theorem}{Theorem}[section]
\newtheorem{prop}[theorem]{Proposition}
\newcommand{\proof}{\noindent{\bf Proof}.\ }

\begin{document}

\title{Structuring quantum effects: superoperators as arrows} 
\author{
Juliana K. Vizzotto${}^1$\thanks{Permanent address: 
  Institute of Informatics, Porto Alegre, Brazil} 
\quad\qquad Thorsten Altenkirch${}^2$ 
\quad\qquad Amr Sabry${}^1$ \\
\ \\
\begin{tabular}{c@{\quad\qquad}c}
${}^1$ Department of Computer Science & 
  ${}^2$ School of Computer Science and IT \\
Indiana University, USA & The University of Nottingham, UK
\end{tabular}
}

\maketitle
\thispagestyle{empty}

\begin{abstract}

We show that the model of quantum computation based on density matrices and
superoperators can be decomposed in a pure classical (functional) part and an
effectful part modeling probabilities and measurement. The effectful part can
be modeled using a generalization of monads called arrows. We express the
resulting executable model of quantum computing in the programming language
Haskell using its special syntax for arrow computations. The embedding in
Haskell is however not perfect: a faithful model of quantum computing
requires type capabilities which are not directly expressible in Haskell.

\end{abstract}

\section{Introduction}

A newcomer to the field of quantum computing is immediately overwhelmed with
many apparent differences with classical computing that suggest that quantum
computing might require radically new semantic models and programming
languages. In some sense this is true for two reasons: (1) quantum computing
is based on a kind of parallelism caused by the non-local wave character of
quantum information which is qualitatively different from the classical
notion of parallelism, and (2) quantum computing has a peculiar notion of
observation in which the observed part of the quantum state and every other
part that is entangled with it immediately lose their wave
character. Interestingly none of the other differences that are often cited
between quantum and classical computing are actually relevant
semantically. For example, even though we do not often think of classical
computation as ``reversible,'' it is just as reversible as quantum
computing. Both can be compiled or explained in terms of reversible
circuits~\cite{alti:qml-draft}, but in neither model should the user be
required to reason about reversibility.

The two properties of quantum computing discussed above certainly go beyond
``pure'' classical programming and it has been suspected earlier that they
might correspond to some notion of computational effect. Following Moggi's
influential paper~\cite{77353}, computational effects like assignments,
exceptions, non-determinism, \textit{etc.} could all be modeled using the
categorical construction of a \emph{monad}. This construction has been
internalized in the programming language Haskell as a tool to elegantly
express computational effects within the context of a pure functional
language. Since the work of Moggi, several natural notions of computational
effects were discovered which could only be expressed as generalizations of
monads. Of particular importance to us, is the generalization of monads known
as arrows~\cite{hughes:arrows} which is also internalized in the programming
language Haskell.

In an early paper, Mu and Bird (2001) showed that quantum parallelism is
almost a monad. We expand and build on this observation as follows. First the
traditional model of quantum computing cannot even express measurements, so
we use a known more general model using density matrices and
superoperators. After expressing this model in Haskell, we establish that the
superoperators used to express all quantum computations and measurements are
indeed an instance of the concept of arrows (with a small caveat). In
particular the construction clarifies the crucial need for some form of
linear typing: arrow computations must be required to use every quantum value
or else the computations produce results that are inconsistent with quantum
mechanics. 

In summary, our construction relates ``exotic'' quantum features to
well-understood semantic constructions and programming languages. We hope it
will serve as a useful tool to further understand the nature and structure of
quantum computation. The remainder of the paper is organized as
follows. Section~\ref{sec:traditional} presents the traditional model of
quantum computing and its implementation in Haskell, focusing on the
possibility of structuring the effects using monads. Section~\ref{sec:meas}
discusses the limitations of the traditional model as a complete model of
quantum computation which should include measurement. Section~\ref{sec:dens}
introduces a more general model of quantum based on density matrices and
superoperators. Our main result is discussed in Section~\ref{sec:arrows}
where we show that general quantum computations including measurement can be
structured using the generalization of monads called
arrows. Section~\ref{sec:examples} gives two complete examples implementing a
Toffoli circuit and the teleportation experiment: both examples use the arrow
notation to express the structure of the computation
elegantly. Section~\ref{sec:qml} discusses the limitations of our model and
its connection to the functional quantum programming language
QML~\cite{alti:qml-draft}.  Section~\ref{sec:conc}
concludes. Appendix~\ref{sec:app1} explains the basics of the Haskell
notation used in the paper, and the next two appendices present the proofs
that are omitted from the main body of the paper.

%%%%%%%%%%%%%%%%%%%%%%%%%%%%%%%%%%%%%%%%%%%%%%%%%%%%%%%%%%%%%%%%%%%%%%%%%%%%%%%%
\section{The Traditional Model of Quantum Computing}
\label{sec:traditional}

We present the traditional model of quantum computing in this section. 

%%%%%%%%%%%%%%%%%%%%%%%%%%%%
\subsection{Vectors}

A finite set $a$ can be represented in Haskell as an instance of the class
\<Basis\> below. Given such a set \<a\> representing observable (classical)
values, a pure quantum value is a vector \<a\to\mathbb{C}\> which associates
each basis element with a complex probability amplitude. The basis elements
must be distinguishable from each other which explains the constraint \<Eq a\> 
on the type of elements below:

\bigskip\begin{tabular}{|c@{\qquad}|}%
\hline%
\begin{minipage}{5in}{
\begin{tabbing}\texfamily
{\bfseries class}~{\itshape Eq}~a~=>~~{\itshape Basis}~a~{\bfseries where}~basis~::~[a]~\\
\texfamily {\bfseries type}~{\itshape PA}~=~{\itshape Complex}~{\itshape Double}\\
\texfamily {\bfseries type}~{\itshape Vec}~a~=~a~\char'31~{\itshape PA}
\end{tabbing}
}\end{minipage}\\%
\hline%
\end{tabular}\bigskip

The type constructor \<Vec\> is technically not a monad: it corresponds to a
\emph{Kleisli structure}~\cite{kleisli}. Yet as noted by Mu and
Bird (2001), the probabilities introduced by vector spaces constitute
a computational effect which can be structured using a slight generalization
of monads in Haskell~\cite{77353}. From a programming perspective, a monad is
represented using a type constructor for computations \<m\> and two
functions: \<return :: a \to m a\> and \<\hsbind :: m a \to (a \to m b) \to m b\>. 
The operation $\hsbind$ (pronounced ``bind'') specifies how to sequence
computations and \<return\> specifies how to terminate computations:

\bigskip\begin{tabular}{|c@{\qquad}|}%
\hline%
\begin{minipage}{5in}{\begin{tabbing}\texfamily
return~::~{\itshape Basis}~a~=>~a~\char'31~{\itshape Vec}~a\\
\texfamily return~a~b~=~{\bfseries if}~a\char'36b~{\bfseries then}~1~{\bfseries else}~0\\
\texfamily \\
\texfamily (>>=)~::~{\itshape Basis}~a~=>~{\itshape Vec}~a~\char'31~(a~\char'31~{\itshape Vec}~b)~\char'31~{\itshape Vec}~b\\
\texfamily va~>>=~f~=~\char'10~b~\char'31~sum~[~(va~a)~*~(f~a~b)~|~a~\char'06~basis]
\end{tabbing}
}\end{minipage}\\%
\hline%
\end{tabular}\bigskip

\noindent Because of the additional constraint that our computations must be
over specified bases whose elements must be comparable, the types of our
operations are more restricted than strictly desired for a monad. However
\<return\> and \<\hsbind\> satisfy the three monad laws.

\begin{prop}
\label{prop:vec}
Vector spaces satisfy the required equations for monads.
\end{prop}
\proof See Appendix~\ref{sec:app2}.\hfill$\Box$

\bigskip

Vector spaces have additional properties abstracted in the Haskell class
\<MonadPlus\>. Instances of this class support two additional methods:
\<mzero\> and \<mplus\> which provide a ``zero'' computation and an operation
to ``add'' computations:

\bigskip\begin{tabular}{|c@{\qquad}|}%
\hline%
\begin{minipage}{5in}{\begin{tabbing}\texfamily
mzero~::~{\itshape Vec}~a~\\
\texfamily mzero~=~const~0~\\
\texfamily \\
\texfamily mplus~::~{\itshape Vec}~a~\char'31~{\itshape Vec}~a~\char'31~{\itshape Vec}~a~\\
\texfamily mplus~v\char95 1~v\char95 2~a~=~v\char95 1~a~+~v\char95 2~a\\
\texfamily \\
\texfamily mminus~::~{\itshape Vec}~a~\char'31~{\itshape Vec}~a~\char'31~{\itshape Vec}~a~\\
\texfamily mminus~v\char95 1~v\char95 2~a~=~v\char95 1~a~-~v\char95 2~a
\end{tabbing}
}\end{minipage}\\%
\hline%
\end{tabular}\bigskip

For convenience, it is also possible to define various kinds of products over
vectors: the \emph{scalar} product~$\scapr$, the \emph{tensor} product
$\tenspr$, and the \emph{dot} product $\dotpr$:

\bigskip\begin{tabular}{|c@{\qquad}|}%
\hline%
\begin{minipage}{5in}{\begin{tabbing}\texfamily
(\$*)~::~{\itshape PA}~\char'31~{\itshape Vec}~a~\char'31~{\itshape Vec}~~a\\
\texfamily pa~\$*~v~=~\char'10a~\char'31~pa~*~v~a\\
\texfamily \\
\texfamily (<*>)~::~{\itshape Vec}~a~\char'31~{\itshape Vec}~b~\char'31~{\itshape Vec}~(a,b)\\
\texfamily v1~<*>~v2~=~\char'10~(a,b)~\char'31~v1~a~*~v2~b\\
\texfamily \\
\texfamily (<.>)~::~{\itshape Basis}~a~=>~{\itshape Vec}~a~\char'31~{\itshape Vec}~a~\char'31~{\itshape PA}\\
\texfamily v1~<.>~v2~=~sum~(map~(\char'10a~\char'31~conjugate~(v1~a)~*~(v2~a))~basis)
\end{tabbing}
}\end{minipage}\\%
\hline%
\end{tabular}\bigskip

Examples of vectors over the set of booleans may be defined as follows:

\begin{tabbing}\texfamily
{\bfseries instance}~{\itshape Basis}~{\itshape Bool}~{\bfseries where}~basis~=~[{\itshape False},{\itshape Bool}]~\\
\texfamily qFalse,qTrue,qFT,qFmT~::~{\itshape Vec}~{\itshape Bool}~\\
\texfamily qFalse~=~return~{\itshape False}\\
\texfamily qTrue~=~return~{\itshape True}\\
\texfamily qFT~=~(1~/~sqrt~2)~\$*~(qFalse~`mplus`~qTrue)\\
\texfamily qFmT~=~(1~/~sqrt~2)~\$*~(qFalse~`mminus`~qTrue)
\end{tabbing}
\noindent The first two are unit vectors corresponding to basis elements; the
last two represent state which are in equal superpositions of \<False\> and
\<True\>. In the Dirac notation, these vectors would be respectively written
as $\dirac{\textit{False}}$, $\dirac{\textit{True}}$, $\frac{1}{\sqrt{2}}
(\dirac{\textit{False}} + \dirac{\textit{True}})$, and $\frac{1}{\sqrt{2}}
(\dirac{\textit{False}} - \dirac{\textit{True}})$.

Vectors over several values can be easily described using the tensor product
on vectors or the Cartesian product on the underlying bases:

\begin{tabbing}\texfamily
{\bfseries instance}~({\itshape Basis}~a,~{\itshape Basis}~b)~=>~{\itshape Basis}(a,~b)~{\bfseries where}~\\
\texfamily basis~=~[(a,~b)~|~a~\char'06~basis,~b~\char'06~basis~]~\\
\texfamily \\
\texfamily p1,p2,p3,epr~::~{\itshape Vec}~({\itshape Bool},{\itshape Bool})~\\
\texfamily \\
\texfamily p1~=~qFT~<*>~qFalse~\\
\texfamily p2~=~qFalse~<*>~qFT~\\
\texfamily p3~=~qFT~<*>~qFT~\\
\texfamily \\
\texfamily epr~({\itshape False},{\itshape False})~=~1~/~sqrt~2~\\
\texfamily epr~({\itshape True},{\itshape True})~=~1~/~sqrt~2
\end{tabbing}

\noindent In contrast to the first three vectors, the last vector describes
an \emph{entangled} quantum state which cannot be separated into the product
of independent quantum states. The name of the vector ``\<epr\>'' refers to
the initials of Einstein, Podolsky, and Rosen who used such a vector in a
thought experiment to demonstrate some strange consequences of quantum
mechanics~\cite{epr}.

%%%%%%%%%%%%%%%%%%%%%%%%%%%%
\subsection{Linear Operators}

Given two base sets $A$ and $B$ a linear operator $f \in A \multimap B$ is a
function mapping vectors over $A$ to vectors over $B$. We represent such
operators as functions mapping values to vectors which is similar to
representation used by Karczmarczuk~(2003):

\bigskip\begin{tabular}{|c@{\qquad}|}%
\hline%
\begin{minipage}{5in}{\begin{tabbing}\texfamily
{\bfseries type}~{\itshape Lin}~a~b~=~a~\char'31~{\itshape Vec}~b~\\
\texfamily \\
\texfamily fun2lin~::~({\itshape Basis}~a,~{\itshape Basis}~b)~=>~(a~\char'31~b)~\char'31~{\itshape Lin}~a~b~\\
\texfamily fun2lin~f~a~=~return~(f~a)
\end{tabbing}
}\end{minipage}\\%
\hline%
\end{tabular}\bigskip

\noindent The function \<fun2lin\> converts a regular function to a linear
operator. For example, the quantum version of the boolean negation is:
\begin{tabbing}\texfamily
qnot~::~{\itshape Lin}~{\itshape Bool}~{\itshape Bool}~\\
\texfamily qnot~=~fun2lin~\char'05
\end{tabbing}
Linear operations can also be defined directly, for example:
\begin{tabbing}\texfamily
phase~::~{\itshape Lin}~{\itshape Bool}~{\itshape Bool}~\\
\texfamily phase~{\itshape False}~=~return~{\itshape False}~\\
\texfamily phase~{\itshape True}~=~(0~:+~1)~\$*~(return~{\itshape True})\\
\texfamily \\
\texfamily hadamard~::~{\itshape Lin}~{\itshape Bool}~{\itshape Bool}~\\
\texfamily hadamard~{\itshape False}~=~qFT~\\
\texfamily hadamard~{\itshape True}~=~qFmT
\end{tabbing}

The definition of a linear operation specifies its action on one individual
element of the basis. To apply a linear operation \<f\> to a vector \<v\>, we
use the \emph{bind} operation to calculate \<v \hsbind f\>. For example
\<(qFT \hsbind hadamard)\> applies the operation \<hadamard\> to the vector
\<qFT\> which one can calculate produces the vector \<qFalse\> as a result.

It is possible to write higher-order functions which consume linear operators
and produce new linear operators. An important example of such functions
produces the so-called \emph{controlled operations}:

\begin{tabbing}\texfamily
controlled~::~{\itshape Basis}~a~=>~{\itshape Lin}~a~a~\char'31~{\itshape Lin}~({\itshape Bool},a)~({\itshape Bool},a)~\\
\texfamily controlled~f~(b1,b2)~=~(return~b1)~<*>~({\bfseries if}~b1~{\bfseries then}~f~b2~{\bfseries else}~return~b2)
\end{tabbing}

\noindent The linear operator \<f\> is transformed to a new linear
operator controlled by a quantum boolean value. The modified operator
returns a pair whose first component is the input control value. The
second input is passed to \<f\> only if the control value is true, and
is otherwise left unchanged. For example, \< (qFT \tenspr qFalse)
\hsbind (controlled qnot) \> applies the familiar
\emph{controlled-not} gate to a vector over two values: the control
value is a superposition of \<False\> and \<True\> and the data value
is \<False\>. As one may calculate the result of this application is
the \<epr\> vector.

Linear operations can be combined and transformed in several ways which we
list below. The function $\outpr$ produces the linear operator corresponding to the
\emph{outer product} of two vectors. The functions \<linplus\> and
\<lintens\> are the functions corresponding to the sum and tensor product on
vectors. Finally the function \<o\> composes two linear operators. 

\bigskip\begin{tabular}{|c@{\qquad}|}%
\hline%
\begin{minipage}{5in}{\begin{tabbing}\texfamily
adjoint~::~{\itshape Lin}~a~b~\char'31~{\itshape Lin}~b~a~\\
\texfamily adjoint~f~b~a~=~conjugate~(f~a~b)~\\
\texfamily \\
\texfamily (>*<)~::~{\itshape Basis}~a~=>~{\itshape Vec}~a~\char'31~{\itshape Vec}~a~\char'31~{\itshape Lin}~a~a~\\
\texfamily (v1~>*<~v2)~a1~a2~=~v1~a1~*~conjugate~(v2~a2)~\\
\texfamily \\
\texfamily linplus~::~({\itshape Basis}~a,~{\itshape Basis}~b)~=>~{\itshape Lin}~a~b~\char'31~{\itshape Lin}~a~b~\char'31~{\itshape Lin}~a~b~\\
\texfamily linplus~f~g~a~=~f~a~`mplus`~g~a~\\
\texfamily \\
\texfamily lintens~::~({\itshape Basis}~a,~{\itshape Basis}~b,~{\itshape Basis}~c,~{\itshape Basis}~d)~=>\\
\texfamily ~~~~~~~~~~~~~~~~{\itshape Lin}~a~b~\char'31~{\itshape Lin}~c~d~\char'31~{\itshape Lin}~(a,c)~(b,d)~\\
\texfamily lintens~f~g~(a,c)~=~f~a~<*>~g~c~\\
\texfamily \\
\texfamily o~::~({\itshape Basis}~a,~{\itshape Basis}~b,~{\itshape Basis}~c)~=>~{\itshape Lin}~a~b~\char'31~{\itshape Lin}~b~c~\char'31~{\itshape Lin}~a~c~\\
\texfamily o~f~g~a~=~(f~a~>>=~g)
\end{tabbing}
}\end{minipage}\\%
\hline%
\end{tabular}\bigskip

%%%%%%%%%%%%%%%%
\subsection{Example: A Toffoli Circuit}
\label{sec:toffoli}

\bigskip
\begin{center}
\scalebox{0.7}{\includegraphics{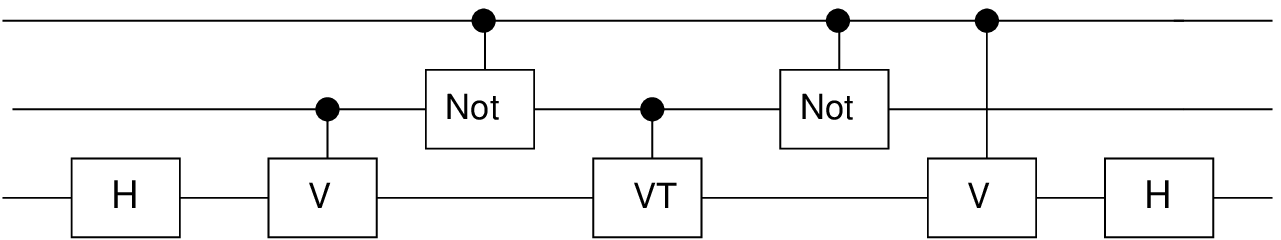}}
\end{center}

\noindent The circuit diagram uses the de-facto standard notation for specifying
quantum computations. Each line carries one quantum bit (\emph{qubit}); we
refer to the three qubits in the circuit as \<top\>, \<middle\>, and
\<bottom\>. The values flow from left to right in steps corresponding to the
alignment of the boxes which represent quantum gates. The gates labeled
\<H\>, \<V\>, \<VT\>, and \<Not\> represent the quantum operations
\<hadamard\>, \<phase\>, \<adjoint phase\>, and \<qnot\> respectively. Gates
connected via a bullet to another wire are \emph{controlled} operations.

In general all three qubits in the circuit may be entangled and hence the
state vector representing them cannot be separated into individual state
vectors. This means that, despite the appearance to the contrary, it is not
possible to operate on any of the lines individually. Instead the circuit
defines a linear operation on the entire state:

\begin{tabbing}\texfamily
toffoli~::~{\itshape Lin}~({\itshape Bool},{\itshape Bool},{\itshape Bool})~({\itshape Bool},{\itshape Bool},{\itshape Bool})~\\
\texfamily toffoli~(top,middle,bottom)~=~\\
\texfamily ~~~{\bfseries let}~cnot~=~controlled~qnot\\
\texfamily ~~~~~~~cphase~=~controlled~phase~\\
\texfamily ~~~~~~~caphase~=~controlled~(adjoint~phase)\\
\texfamily ~~~{\bfseries in}~hadamard~bottom~>>=~\char'10~b1~\char'31\\
\texfamily ~~~~~~cphase~(middle,b1)~>>=~\char'10~(m1,b2)~\char'31~\\
\texfamily ~~~~~~cnot~(top,m1)~>>=~\char'10~(t1,m2)~\char'31~\\
\texfamily ~~~~~~caphase~(m2,b2)~>>=~\char'10~(m3,b3)~\char'31~\\
\texfamily ~~~~~~cnot~(t1,m3)~>>=~\char'10~(t2,m4)~\char'31~\\
\texfamily ~~~~~~cphase~(t2,b3)~>>=~\char'10~(t3,b4)~\char'31~\\
\texfamily ~~~~~~hadamard~b4~>>=~\char'10~b5~\char'31~\\
\texfamily ~~~~~~return~(t3,m4,b5)
\end{tabbing}

%%%%%%%%%%%%%%%%%%%%%%%%%%%%%%%%%%%%%%%%%%%%%%%%%%%%%%%%%%%%%%%%%%%%%%%%%%%%%%%%
\section{Measurement}
\label{sec:meas}

The use of monads to structure the probability effects reveals an elegant
underlying structure for quantum computations. This structure can be studied
in the context of category theory and exploited in the design of a calculus
for quantum
computation~\cite{vanTonder:2003yi,vanTonder:2004,valiron,alti:qml-draft}.

Unfortunately in the traditional model of quantum computing we have
used so far, is difficult or impossible to deal formally with another
class of quantum effects, including measurements, decoherence, or
noise. We first give one example where such effects are critical, and
then discuss various approaches in the literature on how to deal with
such effects.

%%%%%%%%%%%%%%%%%%%
\subsection{Teleportation}
\label{sec:tele}

The idea of teleportation is to disintegrate an object in one place making a
perfect replica of it somewhere else. Indeed quantum
teleportation~\cite{bennett93teleporting} enables the transmission,
\emph{using a classical communication channel}, of an unknown quantum state
via a previously shared \<epr\> pair. 

In the following diagram, Alice and Bob initially have access to one
of the qubits of an entangled \<epr\> pair, and Alice aims to teleport
an unknown qubit $q$ to Bob:

\bigskip 
\begin{center}
\scalebox{0.7}{\includegraphics{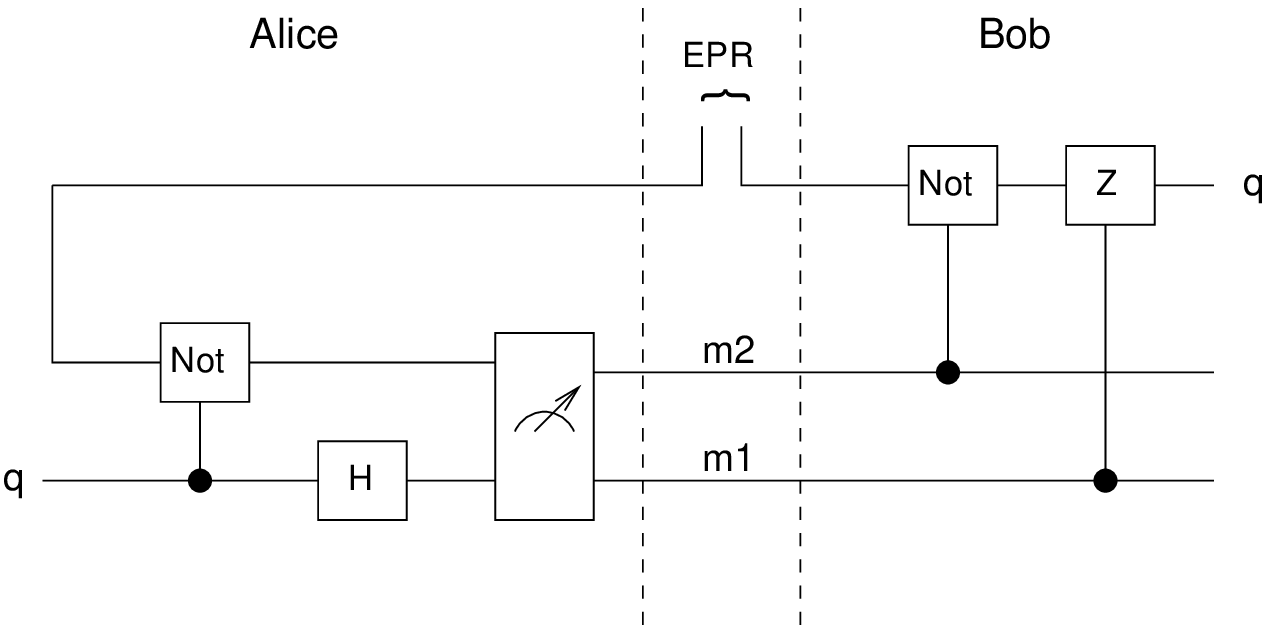}}
\end{center}

The calculation proceeds as follows. First Alice interacts with the unknown
qubit $q$ and her half of the \<epr\> state. Then Alice performs a measurement
collapsing her quantum state and getting two classical bits $m_1$ and $m_2$
that she transmits to Bob using a classical channel of communication. 

Upon receiving the two classical bits of information, Bob interacts with his
half of the \<epr\> state with gates controlled by the classical bits. The
circuit in the figure can be shown to re-create the quantum state~$q$ which
existed at Alice's site before the experiment.

Our main interest in this circuit is that it is naturally expressed
using a sequence of operations on quantum values which include a
non-unitary \emph{measurement} in the middle. Using the model
developed in the previous section, it is not possible to describe this
algorithm as stated. In the next section, we briefly several possible
ways to deal with this problem.

%%%%%%%%%%%%%%%%%%
\subsection{Dealing with Measurement}

The literature includes several approaches to the problem of measurement. We
characterize such approaches in three broad categories: deferring
measurements, using classical control with pointers and side-effects, and
using density matrices and superoperators. We discuss the first two
approaches in the remainder of this section, and expand on the latter
approach in the next section.

%%%%%%%%%%%%%%
\subsubsection{Deferring measurements: }

The first approach (used for example by Mu and Bird (2001), Van Tonder~(2003;
2004) and Karczmarczuk~(2003) relies on the \textit{principle of deferred
measurement}~\cite{NielsenChuang00}. This principle can be used to transform
computations to \textit{always} defer measurements to the end. Using this
idea one can focus entirely on managing the probability effects and perform
the measurements outside the formalism. The drawback of this approach is
clear: programs that interleave quantum operations with measurements cannot
be expressed naturally. For example, transforming the teleportation circuit
above to defer the measurements until after Bob's computation completely
changes the character of the experiment, because no \textit{classical}
information is transmitted form Alice to Bob.

%%%%%%%%%%%%%%
\subsubsection{Classical Control and Side-effects: }

In general, this category of models is based on the so-called QRAM (quantum
random access machine) model of Knill~(1996), which is summarized by the
slogan ``quantum data, classical control''~\cite{selinger}.  In this context,
a quantum computer can be seen as a classical computer with a quantum device
attached to it. The classical control sends instructions for the quantum
machine to execute unitary operations and measurements.  A measurement
collapses the quantum (probabilistic) computation and forces it to produce a
classical (deterministic) result. In fact, the situation is even more
complicated: measuring part of a quantum state collapses not only the
measured part but any other part of the global state with which it is
entangled. The most common approach to computationally realize this hybrid
architecture is via manipulating what are effectively \emph{pointers} to a
\emph{global shared quantum state} as the following examples show:

\begin{itemize} \item In the flowchart notation for the language introduced
by Selinger~(2004), the state is represented by a collection of variables
that can each be assigned \emph{once}. An operation can only be applied to an
initial group of the variables (and is implicitly composed with the identity
on the remaining variables). If the variables are not in the desired order,
they must be permuted first. Thus the first few steps of the \<toffoli\>
circuit are: \begin{center}
\scalebox{0.6}{\includegraphics{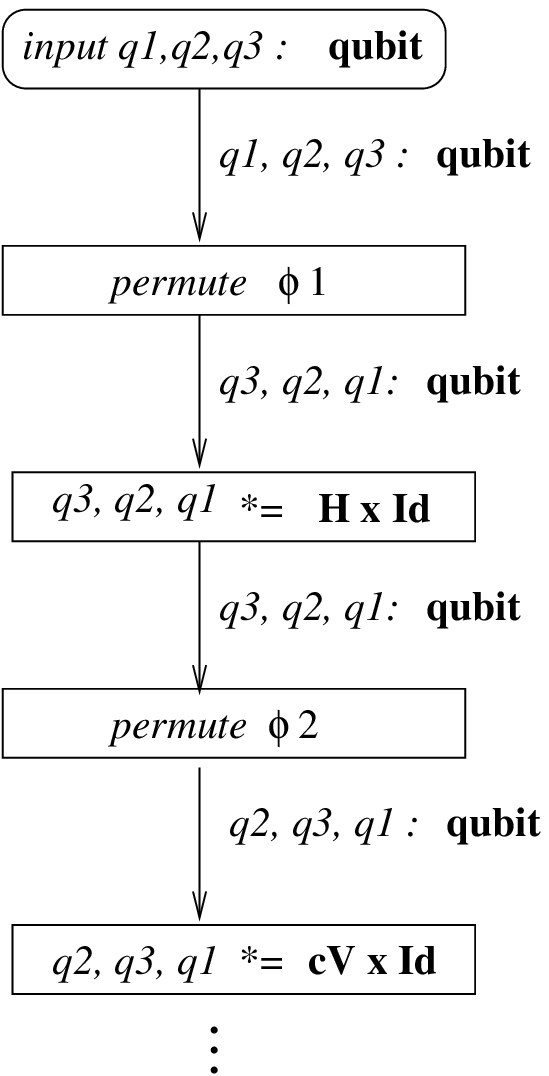}} \end{center}

\item In the procedural language QCL~\cite{qcl} a \emph{quantum register} is
a realized using \emph{pointers} to the complete state. Operations on a
register map to operations on the state as follows. If we have an $m$-qubit
register $r$ which points to an $n$-qubit state, then an operation~$U$ on the
register is realized using:
\[
U(r) = \Pi_r^{\dagger}~(U \times I(n-m))~\Pi_r
\]
The operation $U$ is composed with the identity on the remaining number of
qubits of the state. The operator $\Pi_r$ is an arbitrary reordering operator
and $\Pi_r^{\dagger}$ is its inverse. After re-ordering, the lifted $U$
composed with the identity is applied, and the result is permuted back to the
original order.

\item Jan {Skibi\'nski}~(2001) produced an early Haskell simulator of a
quantum computer. The simulator maintains quantum registers and allows
operations to act on specific qubits using what is essentially pointers. To
apply an operation to the third, fifth, and seventh qubits on a quantum
register, some low-level calculations depending on the indices and size of
the register are used to produce a lifted operation composed with several
identity operations that acts on the entire register.

\item Valiron~\textit{et. al.} (2004) develop a functional quantum
programming language based on the original work of Selinger~(2004). The
representation of quantum data in their calculus uses an external \<n\>-qubit
state \<Q\>. Programs may contain free variables which are essentially
pointers to the quantum state.

\item In our previous work~\cite{871900} we introduced \textit{virtual
values} to hide the management of pointers to the global state. Using
virtual values the code for the \<toffoli\> example is essentially
identical to the one presented earlier, \emph{except} for the need to
manually generate the \emph{adaptors} which mediate between the
virtual value and the global state.

\end{itemize}

The use of pointers and sharing to model the side-effect of
measurement is in some sense adequate. However by doing so, we
completely lose the monadic structure and the direct connections to
categorical semantics.

%%%%%%%%%%%%%%%%%%%%%%%%%%%%%%%%%%%%%%%%%%%%%%%%%%%%%%%%%%%%%%%%%%%%%%%%%%%%%%%%
\section{Density Matrices and Superoperators}
\label{sec:dens}

Fortunately the usual model of quantum computing can be generalized to
solve the problem of modeling measurements in a better way. In the
generalized model, the state of the computation is represented using a
\emph{density matrix} and the operations are represented using
\emph{superoperators}~\cite{276708}. Using these notions, the
\emph{projections} necessary to express measurements become
expressible within the model. We review this model in this section. 

%%%%%%%%%%%%%%%%%%%%%%%%%%%%%%%

\subsection{Density Matrices}

Intuitively, density matrices can be understood as a statistical perspective of the 
state vector. In the density matrix formalism, a quantum state that used to be
modeled by a vector $v$ is now modeled by its outer product. 

\bigskip\begin{tabular}{|c@{\qquad}|}%
\hline%
\begin{minipage}{5in}{\begin{tabbing}\texfamily
{\bfseries type}~{\itshape Dens}~a~=~{\itshape Vec}~(a,a)\\
\texfamily \\
\texfamily pureD~::~{\itshape Basis}~a~=>~{\itshape Vec}~a~\char'31~{\itshape Dens}~a~\\
\texfamily pureD~v~=~lin2vec~(v~>*<~v)~\\
\texfamily \\
\texfamily lin2vec~::~(a~\char'31~{\itshape Vec}~b)~\char'31~{\itshape Vec}~(a,b)\\
\texfamily lin2vec~=~uncurry
\end{tabbing}
}\end{minipage}\\%
\hline%
\end{tabular}\bigskip

The function \<pureD\> embeds a state vector in its density matrix
representation. For convenience, we uncurry the arguments to the
density matrix so that it looks more like a ``matrix.'' For example,
the density matrices corresponding to the vectors \<qFalse\>,
\<qTrue\>, and \<qFT\> can be visually represented as follows:
\[
\left(\begin{array}{cc}
  1 & 0 \\
  0 & 0 
\end{array}\right)
\qquad\qquad
\left(\begin{array}{cc}
  0 & 0 \\
  0 & 1 
\end{array}\right)
\qquad\qquad
\left(\begin{array}{cc}
  1/2 & 1/2 \\
  1/2 & 1/2 
\end{array}\right)
\]

The appeal of density matrices is that they can represent states other
than the pure ones above. In particular if we perform a measurement on
the state represented by \<qFT\>, we should get \<False\> with
probability $1/2$ or \<True\> with probability $1/2$. This information
which cannot be expressed using vectors, can be represented by the
following density matrix:
\[
\left(\begin{array}{cc}
  1/2 & 0 \\
  0 & 0\\
\end{array}\right)
+ 
\left(\begin{array}{cc}
  0 & 0 \\
  0 & 1/2 \\
\end{array}\right) =
\left(\begin{array}{cc}
  1/2 & 0 \\
  0 & 1/2 \\
\end{array}\right)
\]

Such a density matrix represents a \emph{mixed state} which corresponds to
the sum (and then normalization) of the density matrices for the two results
of the observation. If we further calculate with the result of measuring
\<qFT\> by for example, applying the \<hadamard\> operation, we get one of
the two vectors \<qFT\> or \<qFmT\>, each with probability $1/2$. Because all
operations on vectors are \emph{linear}, we can express this step as follows:

\[
H \;\left(\begin{array}{cc}
  1/2 & 0 \\
  0 & 1/2 \\
\end{array}\right) =
H \; \left(\begin{array}{cc}
  1/2 & 0 \\
  0 & 0\\
\end{array}\right)
+ 
H \; \left(\begin{array}{cc}
  0 & 0 \\
  0 & 1/2 \\
\end{array}\right) =
\left(\begin{array}{cc}
  1/2 & 0 \\
  0 & 1/2 \\
\end{array}\right)
\]

\noindent As the calculation shows, the application of \<hadamard\> has no
effect on the density matrix, and indeed there is no observable difference
between the two configurations before and after the application of
\<hadamard\>. Indeed, the density matrix representation loses the information
in the state vectors that is not observable~\cite{selinger} and hence is a
better representation from a semantic perspective. 

%%%%%%%%%%%%%%%%%%%%%%%%%%%%%%%
\subsection{Superoperators}

Operations mapping density matrices to density matrices are called
\emph{superoperators}: 

\bigskip\begin{tabular}{|c@{\qquad}|}%
\hline%
\begin{minipage}{5in}{\begin{tabbing}\texfamily
{\bfseries type}~{\itshape Super}~a~b~=~(a,a)~\char'31~{\itshape Dens}~b~\\
\texfamily \\
\texfamily lin2super~::~({\itshape Basis}~a,~{\itshape Basis}~b)~=>~{\itshape Lin}~a~b~\char'31~{\itshape Super}~a~b\\
\texfamily lin2super~f~(a1,a2)~=~(f~a1)~<*>~(dual~(adjoint~f)~a2)\\
\texfamily ~~~~~{\bfseries where}~dual~f~a~b~=~f~b~a
\end{tabbing}
}\end{minipage}\\%
\hline%
\end{tabular}\bigskip

\noindent The function \<lin2super\> constructs a superoperator from a
linear operator on vectors. To understand the basic idea, consider the
density matrix resulting from the application of $f$ to
$\dirac{v}$. This corresponds to the outer product of the vector
$f\dirac{v}$ with itself, which applies $f$ to $\dirac{v}$ and the
adjoint of $f$ to the ``dual vector.''

%%%%%%%%%%%%%%%%
\subsection{Tracing and Measurement}

In contrast to the situation with the traditional model of quantum
computing, it is possible to define a superoperator which ``forgets'',
\emph{projects}, or \emph{traces out} part of a quantum state as well
as a superoperator which \emph{measures} part of a quantum state:

\bigskip\begin{tabular}{|c@{\qquad}|}%
\hline%
\begin{minipage}{5in}{\begin{tabbing}\texfamily
trL~::~({\itshape Basis}~a,~{\itshape Basis}~b)~=>~{\itshape Super}~(a,b)~b~\\
\texfamily trL~((a1,b1),(a2,b2))~=~~{\bfseries if}~a1~\char'36~a2~{\bfseries then}~return~(b1,b2)~{\bfseries else}~mzero~\\
\texfamily \\
\texfamily meas~::~{\itshape Basis}~a~=>~{\itshape Super}~a~(a,a)~\\
\texfamily meas~(a1,a2)~=~{\bfseries if}~a1~\char'36~a2~{\bfseries then}~return~((a1,a1),(a1,a1))~{\bfseries else}~mzero
\end{tabbing}
}\end{minipage}\\%
\hline%
\end{tabular}\bigskip

For example, the sequence:

\begin{tabbing}\texfamily
pureD~qFT~>>=~meas~>>=~trL
\end{tabbing}

\noindent first performs a measurement on the pure density matrix
representing the vector \<qFT\>. This measurement produces a vector
with two components: the first is the resulting collapsed quantum
state and the second is the classical observed value. The last
operation forgets about the collapsed quantum state and returns the
result of the classical measurement. As explained earlier the
resulting density matrix is: 
\[
\left(\begin{array}{cc}
  1/2 & 0 \\
  0 & 1/2 \\
\end{array}\right)
\]

%%%%%%%%%%%%%%%%%%%%%%%%%%%%%%%%%%%%%%%%%%%%%%%%%%%%%%%%%%%%%%%%%%%%%%%%%%%%%%%%
\section{Superoperators as Arrows}
\label{sec:arrows}

By moving to density matrices and superoperators, it becomes possible to
express both the original computations as well as measurements in the same
formalism. One might hope that the original monadic structure of quantum
computations is preserved, but it appears that this is not the case.  The
best we can do is to prove that the new model of computation fits within a
generalization of monads called \emph{arrows}.

%%%%%%%%%%%%%%%%%%%%%%%%
\subsection{Arrows}

The application of a superoperator to a density matrix can still be achieved
with the monadic bind operation, instantiated to the following type:

\begin{tabbing}\texfamily
>>=~::~{\itshape Dens}~a~\char'31~((a,a)~\char'31~{\itshape Dens}~b)~\char'31~{\itshape Dens}~b
\end{tabbing}

This type does not however correspond to the required type as computations
now \emph{consume multiple input values}. This observation is reminiscent of
Hughes's motivation for generalizing monads to
\emph{arrows}~\cite{hughes:arrows}.  Indeed, in addition to defining a notion
of procedure which may perform computational effects, arrows may have a
static component independent of the input, or may accept more than one input.

In Haskell, the arrow interface is defined using the following class
declaration:
\begin{tabbing}\texfamily
{\bfseries class}~{\itshape Arrow}~a~{\bfseries where}\\
\texfamily ~~arr~::~~(b~\char'31~c)~\char'31~a~b~c~\\
\texfamily ~~(>>>)~~::~~a~b~c~\char'31~a~c~d~\char'31~a~b~d\\
\texfamily ~~first~~::~~a~b~c~\char'31~a~(b,d)~(c,d)
\end{tabbing}

\noindent In other words, to be an arrow, a type $a$ must support the
three operations $arr$, $\ggg$, and $\textit{first}$ with the given
types. The operations must satisfy the following equations:

\[
\begin{array}{rcl}\hline
\textit{arr} \; \textit{id} \ggg f &=& f\\
f \ggg \textit{arr} \; \textit{id} &=& f\\
(f \ggg g) \ggg h &=& f \ggg (g \ggg h)\\
\textit{arr}\; (g \;.\; f) &=& \textit{arr} \; f \ggg \textit{arr} \; g\\
\textit{first}\;(\textit{arr} \; f) &=& \textit{arr} \;(f \times \textit{id})\\
\textit{first} \;(f \ggg g) &=& \textit{first} \; f \ggg \textit{first} \; g\\
\textit{first} \; f \ggg \textit{arr} \;(\textit{id} \times g) &=& 
  \textit{arr}\; (\textit{id} \times g) \ggg \textit{first} \; f\\
\textit{first} \; f \ggg \textit{arr} \; \textit{fst} &=& 
  \textit{arr} \; \textit{fst} \ggg f\\
\textit{first} \; (\textit{first} \; f) \ggg \textit{arr} \; \textit{assoc} &=& 
  \textit{arr} \; \textit{assoc} \ggg \textit{first} \; f\\
\hline
\end{array}
\]
where the functions $\times$ and $\textit{assoc}$ are defined as follows:

\[
\begin{array}{l}
(f \times g) \;(a,b) = (f \;a, g \;b) \\
assoc \;((a,b),c) = (a,(b,c))
\end{array}
\]

Graphically the functions associated with the arrow type are the following:

\begin{center}
\scalebox{0.9}{\includegraphics{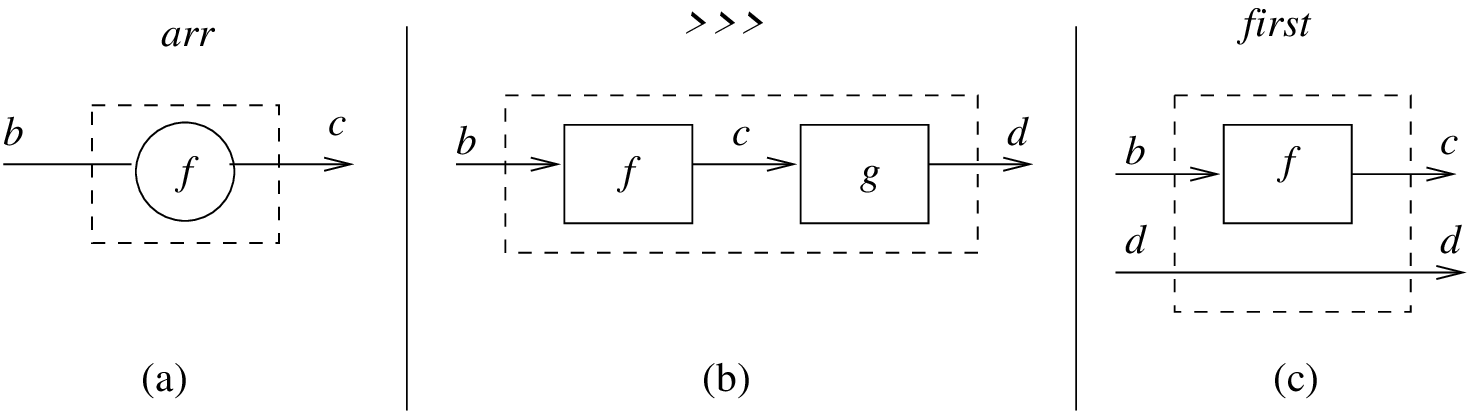}}
\end{center}

The function \textit{arr} allows us to introduce ``pure'' arrows which are
simple functions from their inputs to their outputs. The function $\ggg$ is
similar to $\hsbind$: it composes two computations. The function
\textit{first} is the critical one for our purposes: it allows us to apply an
arrow to a component of the global quantum state. The equations above ensure
that these operations are always well-defined even with arbitrary
permutations and change of associativity.

%%%%%%%%%%%%%%%%%%%%%%%%%%%%%%%%%%%%%%%%%%%
\subsection{Superoperators are Arrows (with Eq constraint)}

Just as the probability effect associated with vectors is not strictly a
monad because of the \<Basis\> constraint, the type \<Super\> is not strictly
an arrow as the following types include the additional constraint requiring
the elements to be comparable:

\begin{tabbing}\texfamily
arr~::~({\itshape Basis}~b,~{\itshape Basis}~c)~=>~(b~\char'31~c)~\char'31~{\itshape Super}~b~c~\\
\texfamily arr~f~=~fun2lin~(\char'10~(b1,b2)~\char'31~(f~b1,~f~b2))~\\
\texfamily \\
\texfamily (>>>)~::~({\itshape Basis}~b,~{\itshape Basis}~c,~{\itshape Basis}~d)~=>~\\
\texfamily ~~~~~~~~~~~~~~~{\itshape Super}~b~c~\char'31~{\itshape Super}~c~d~\char'31~{\itshape Super}~b~d~\\
\texfamily (>>>)~=~o~\\
\texfamily \\
\texfamily first~::~({\itshape Basis}~b,~{\itshape Basis}~c,~{\itshape Basis}~d)~=>~{\itshape Super}~b~c~\char'31~{\itshape Super}~(b,d)~(c,d)~\\
\texfamily first~f~((b1,d1),(b2,d2))~=~permute~((f~(b1,b2))~<*>~(return~(d1,d2)))~\\
\texfamily ~~~{\bfseries where}~permute~v~((b1,b2),(d1,d2))~=~v~((b1,d1),(b2,d2))
\end{tabbing}
\noindent The function \textit{arr} constructs a superoperator from a
pure function by applying the function to both the vector and its
dual. The composition of arrows is simply the composition of linear
operators. The function \textit{first} applies the superoperator $f$
to the first component (and its dual) and leaves the second component
unchanged. The definition calculates each part separately and then
permutes the results to match the required type.

\begin{prop}
\label{prop:super}
Superoperators satisfy the required equations for arrows. 
\end{prop}
\proof See Appendix \ref{sec:app3}.\hfill$\Box$

\bigskip

The proposition implies that we can use the arrow combinators to
structure our computations. For instance, the first few steps of the
Toffoli circuit of Section~\ref{sec:toffoli} would now look like:

\begin{tabbing}\texfamily
toffoli~::~{\itshape Super}~({\itshape Bool},{\itshape Bool},{\itshape Bool})~({\itshape Bool},{\itshape Bool},{\itshape Bool})~\\
\texfamily toffoli~=~\\
\texfamily ~~{\bfseries let}~hadS~=~lin2super~hadamard~\\
\texfamily ~~~~~~cphaseS~=~lin2super~(controlled~phase)\\
\texfamily ~~~~~~cnotS~=~lin2super~(controlled~qnot)\\
\texfamily ~~{\bfseries in}~arr~(\char'10~(a0,~b0,~c0)~\char'31~(c0,~(a0,~b0)))~>>>\\
\texfamily ~~~~~(first~hadS~~>>>~arr~(\char'10~(c1,~(a0,~b0))~\char'31~((b0,~c1),~a0)))~>>>\\
\texfamily ~~~~~(first~cphaseS~>>>~arr~(\char'10~((b1,~c2),~a0)~\char'31~((a0,~b1),~c2)))~>>>\\
\texfamily ~~~~~(first~cnotS~>>>~arr~(\char'10~((a1,~b2),~c2)~\char'31~((b2,~c2),~a1)))~>>>~...
\end{tabbing}

Clearly this notation is awkward as it forces us to explicitly
manipulate the entire state and to manually permute the
values. However, all the tedious code can be generated automatically
as we explain next.

%%%%%%%%%%%%%%%%%%%%%%%%%%%%%%%%%%%%%%%%%%%%%%%%%%%%%%%%%%%%
\subsection{A Better Notation for Arrows}

Following the Haskell's monadic \<\hskwd{do}\>-notation,
Paterson~(2001) presented an extension to Haskell
with an improved syntax for writing computations using arrows. We
concentrate only on the explanation of new forms which we use in our
examples.  Here is a simple example to illustrate the notation:

\begin{tabbing}\texfamily
e1~::~{\itshape Super}~({\itshape Bool},a)~({\itshape Bool},a)~\\
\texfamily e1~=~proc~(a,b)~\char'31~{\bfseries do}~\\
\texfamily ~~~~~~~~~~r~$\leftarrow$~lin2super~hadamard~$\prec$~a\\
\texfamily ~~~~~~~~~~returnA~$\prec$~(r,b)
\end{tabbing}

\noindent The \<\hskwd{do}\>-notation simply sequences the actions in its
body. The function \<returnA\> is the equivalent for arrows of the
monadic function \<return\>. The two additional keywords are:
\begin{itemize}
\item the \emph{arrow abstraction} \<\hskwd{proc}\> which constructs
an arrow instead of a regular function. 
\item the \emph{arrow application} $\prec$ which feeds the value
of an expression into an arrow.
\end{itemize}

Paterson~(2001) shows that the above notation is
general enough to express arrow computations and implemented a
preprocessor which translates the new syntax to regular Haskell. In
the case of \texttt{e1} above, the translation to Haskell produces the following code:
\begin{tabbing}\texfamily
e2~::~{\itshape Super}~({\itshape Bool},a)~({\itshape Bool},a)\\
\texfamily e2~=~first~(lin2super~hadamard)
\end{tabbing}
As the example shows, the output of the preprocessor is quite
optimized. 

%%%%%%%%%%%%%%%%%%%%%%%%%%%%%
\subsection{Superoperators are (probably) not monads}
\label{sec:arrowApp}

Arrows are more general than monads. In particular, they include notions of
computation that consume multiple inputs as well as computations with static
components, independent of the input. Due to this general aspect of arrows,
there are some subclasses of them which turns out to be equivalent to
monads. More precisely, arrow types which support the following \<app\>
function are just as expressive as monads.

\begin{tabbing}\texfamily
{\bfseries class}~{\itshape Arrow}~=>~{\itshape ArrowApply}~a~{\bfseries where}\\
\texfamily ~~app~::~a~(a~b~c,~b)~c
\end{tabbing}

\noindent In other words, for superoperators to be monads, we would have to
define a superoperator of type:
\begin{tabbing}\texfamily
{\itshape Super}~({\itshape Super}~b~c,~b)~c
\end{tabbing}

\noindent which in our case would require \<Super b c\> to be an instance of
\<Basis\>.  Unfortunately there is no straightforward way to view the space
of superoperators as a finite set of observables.

\omitnow{ First, because in general superoperators do not form a basis and
\<app\> is a higher-order arrow, they fail as an instance of
\<ArrowApply\>. Also, the combinator \<app\> describes computations that
always take a single input, which is not the case for superoperators
(\textit{i.e.}, \< Super a b = (a,a) \to Dens b\>). The \<app\> function
gives the directives to define the monadic combinator \<\hsbind\>.  }

%%%%%%%%%%%%%%%%%%%%%%%%%%%%%%%%%%%%%%%%%%%%%%%%%%%%%%%%%%%%%%%%%%%%%%%%%%%%%%%%
\section{Examples Revisited: Toffoli and Teleportation}
\label{sec:examples}

Using arrows and the notation introduced by Patterson, we can express
both of our examples elegantly. 

%%%%%%%%%%
\subsection{Toffoli}

The code mirrors the structure of the circuit and the structure of the
monadic computation expressed earlier:

\begin{tabbing}\texfamily
toffoli~::~{\itshape Super}~({\itshape Bool},{\itshape Bool},{\itshape Bool})~({\itshape Bool},{\itshape Bool},{\itshape Bool})\\
\texfamily toffoli~=~{\bfseries let}~hadS~=~lin2super~hadamard~\\
\texfamily ~~~~~~~~~~~~~~cnotS~=~lin2super~(controlled~qnot)~\\
\texfamily ~~~~~~~~~~~~~~cphaseS~=~lin2super~(controlled~phase)\\
\texfamily ~~~~~~~~~~~~~~caphaseS~=~lin2super~(controlled~(adjoint~phase))\\
\texfamily ~~~~~~~~~~{\bfseries in}~proc~(a0,b0,c0)~\char'31~{\bfseries do}\\
\texfamily ~~~~~~~~~~~~~c1~$\leftarrow$~hadS~$\prec$~c0\\
\texfamily ~~~~~~~~~~~~~(b1,c2)~$\leftarrow$~cphaseS~$\prec$~(b0,c1)\\
\texfamily ~~~~~~~~~~~~~(a1,b2)~$\leftarrow$~cnotS~$\prec$~(a0,b1)\\
\texfamily ~~~~~~~~~~~~~(b3,c3)~$\leftarrow$~caphaseS~$\prec$~(b2,c2)\\
\texfamily ~~~~~~~~~~~~~(a2,b4)~$\leftarrow$~cnotS~$\prec$~(a1,b3)\\
\texfamily ~~~~~~~~~~~~~(a3,c4)~$\leftarrow$~cphaseS~$\prec$~(a2,c3)\\
\texfamily ~~~~~~~~~~~~~c5~$\leftarrow$~hadS~$\prec$~c4\\
\texfamily ~~~~~~~~~~~~~returnA~$\prec$~(a3,b4,c5)
\end{tabbing}

%%%%%%%%%%%%%%
\subsection{Teleportation}

We use the machinery we have developed to faithfully express the
circuit presented in Section~\ref{sec:tele}. We break the algorithm in
two individual procedures, \<alice\> and \<bob\>.  Besides the use of
the arrows notation to express the action of superoperators on
specific qubits, we incorporate the measurement in Alice's procedure,
and trace out the irrelevant qubits from the answer returned by
Bob. 

\begin{tabbing}\texfamily
alice~::~{\itshape Super}~({\itshape Bool},{\itshape Bool})~({\itshape Bool},{\itshape Bool})~\\
\texfamily alice~=~proc~(eprL,q)~\char'31~{\bfseries do}~\\
\texfamily ~~~~~~~~~~~~~(q1,e1)~$\leftarrow$~(lin2super~(controlled~qnot))~$\prec$~(q,eprL)~\\
\texfamily ~~~~~~~~~~~~~q2~$\leftarrow$~(lin2super~hadamard)~$\prec$~q1~\\
\texfamily ~~~~~~~~~~~~~((q3,e2),(m1,m2))~$\leftarrow$~meas~$\prec$~(q2,e1)~\\
\texfamily ~~~~~~~~~~~~~(m1',m2')~$\leftarrow$~trL~((q3,e2),(m1,m2))~~~\\
\texfamily ~~~~~~~~~~~~~returnA~$\prec$~(m1',m2')\\
\texfamily \\
\texfamily bob~::~{\itshape Super}~({\itshape Bool},{\itshape Bool},{\itshape Bool})~{\itshape Bool}~\\
\texfamily bob~=~proc~(eprR,m1,m2)~\char'31~{\bfseries do}~\\
\texfamily ~~~~~~~~~~~(m2',e1)~$\leftarrow$~(lin2super~(controlled~qnot))~$\prec$~(m2,eprR)~\\
\texfamily ~~~~~~~~~~~(m1',e2)~$\leftarrow$~(lin2super~(controlled~z))~$\prec$~(m1,e1)~\\
\texfamily ~~~~~~~~~~~q'~$\leftarrow$~trL~$\prec$~((m1',m2'),e2)~\\
\texfamily ~~~~~~~~~~~returnA~$\prec$~q'~\\
\texfamily ~\\
\texfamily teleport~::~{\itshape Super}~({\itshape Bool},{\itshape Bool},{\itshape Bool})~{\itshape Bool}~\\
\texfamily teleport~=~proc~(eprL,eprR,q)~\char'31~{\bfseries do}\\
\texfamily ~~~~~~~~~~~~~~~~(m1,m2)~$\leftarrow$~alice~$\prec$~(eprL,q)~\\
\texfamily ~~~~~~~~~~~~~~~~q'~$\leftarrow$~bob~$\prec$~(eprR,m1,m2)~\\
\texfamily ~~~~~~~~~~~~~~~~returnA~$\prec$~q'
\end{tabbing}

%%%%%%%%%%%%%%%%%%%%%%%%%%%%%%%%%%%%%%%%%%%%%%%%%%%%%%%%%%%%%%%%%%%%%%%%%%%%%%%%
\section{Linear Typing: QML}
\label{sec:qml}

The category of superoperators is considered to be an adequate model of
non-reversible quantum computation \cite{selinger}. Our construction
presented so far seems to suggest that this category corresponds to a
functional language with arrows, and so that we can accurately express
quantum computation in such a framework. But as we explain below, this is not
quite the whole story. 

First consider the well-known ``non-cloning'' property of quantum
states~\cite{NielsenChuang00}. The arrow notation allows us to reuse 
variables more than once, and we are free to define the following 
operator:
 
\begin{tabbing}\texfamily
copy~::~{\itshape Super}~{\itshape Bool}~({\itshape Bool},~{\itshape Bool})~\\
\texfamily copy~=~arr~(\char'10~x~\char'31~(x,x))
\end{tabbing}

\noindent But can this superoperator be used to clone a qubit?  The answer,
as explained in Section 1.3.5 of the classic book on quantum
computing~\cite{NielsenChuang00}, is no. The superoperator \<copy\> can be
used to copy classical information encoded in quantum data, but when applied
to an arbitrary quantum state, for example like \<qFT\>, the superoperator
does not make two copies of the state \<qFT\> but rather it produces the
\<epr\> state which is the correct and desired behavior. Thus, in this aspect
the semantics of arrows is coherent with quantum computation, \textit{i.e.},
the use of variables more than once models sharing, not cloning.

In contrast, in our model there is nothing to prevent the definition 
of:

\begin{tabbing}\texfamily
weaken~::~{\itshape Super}~({\itshape Bool},{\itshape Bool})~{\itshape Bool}~\\
\texfamily weaken~=~arr~(\char'10~(x,y)~\char'31~y)
\end{tabbing}

\noindent This operator is however not physically realizable. Applying
\<weaken\> to \<epr\> gives \<qFT\>. Physically forgetting about \<x\>
corresponds to a measurement: if we measure the left qubit of \<epr\> we
should get \<qFalse\> or \<qTrue\> or the mixed state of both measurements,
but never \<qFT\>.

Therefore, our use of Haskell as a vehicle for expressing the ideas finally
hits a major obstacle: arrow computations must be required to use every value
that is introduced. Instead of attempting to continue working within Haskell,
a better approach might be to now consider a functional quantum language like
QML whose type system is designed to explicitly control weakening and
decoherence, and to express the separation of values and arrow computations
in that framework.

In more detail, QML \cite{alti:qml-draft} is a functional quantum programming
language which addresses this problem by using a type system based on strict
linear logic: contraction is permitted and modelled by \<copy\> while
weakening has to be explicit and is translated by a partial trace.  QML also
features to case operators: a classical case operator which measures a qbit
and returns the appropriate branch and a quantum case operator which avoids
measurement but requires that the branches return results in orthogonal
subspaces.

QML programs can be compiled to quantum circuits, using the category of
finite quantum computation FQC --- Grattage's QML compiler
\cite{alti:qmlc-draft} is based on this semantics. An irreversible
computation can be modelled by a reversible circuit, allowing additional heap
qubits, which are initialized to a predefined values at the beginning of the
computation and disposing, \textit{i.e.} measuring, qbits at the end of the
computation.  To any FQC morphism we can assign a superoperator and indeed
every superoperator can be represented this way.

Alternatively, we can interpret QML programs directly as superoperators,
giving rise to a constructive denotational semantics exploiting the 
library of arrow combinators developed here. We hope to exploit this
semantics to further analyze QML and to develop high level reasoning
principles for QML programs.

% ArrowChoice

%%%%%%%%%%%%%%%%%%%%%%%%%%%%%%%%%%%%%%%%%%%%%%%%%%%%%%%%%%%%%%%%%%%%%%%%%%%%%%%%
\section{Conclusion}
\label{sec:conc}

%% Need to talk about QML and quantum control

We have argued that a realistic model for quantum computations should
accommodate both unitary operations and measurements, and we have shown that
such \textit{general} quantum computations can be modeled using arrows. This
is an extension of the previous-known observation that one can model pure
quantum probabilities using monads. Establishing such connections between
quantum computations and monads and arrows enables elegant embeddings in
current classical languages, and exposes connections to well-understood
concepts from the semantics of (classical) programming languages. We have
demonstrated the use of arrows to model elegantly two examples in Haskell,
including the teleportation experiment which interleaves measurements with
unitary operations.

\section*{Acknowledgments}

We would like to thank Ant\^onio Carlos da Rocha Costa and Jonathan Grattage
for extensive discussions and feedback.

%%%%%%%%%%%%%%%%%%%%%%%%%%%%%%%%%%%%%%%%%%%%%%%%%%%%%%%%%%%%%%%%%%%%%%%%%%%%%%%%
\appendix
\section{A Haskell Primer}
\label{sec:app1}

We use Haskell as a precise mathematical (and executable) notation. 

It is useful to think of a Haskell type as representing a mathematical
set. Haskell includes several built-in types that we use: the type
\<Boolean\> whose only two elements are \<False\> and \<True\>; the type
\<Complex Double\> whose elements are complex numbers written $\co{a}{b}$
where both~$a$ and $b$ are elements of the type \<Double\> which approximates
the real numbers. Given two types $a$ and $b$, the type $(a,b)$ is the type
of ordered pairs whose elements are of the respective types; the type $a\to
b$ is the type of functions mapping elements of $a$ to elements of $b$; and
the type $[a]$ is the type of sequences (lists) whose elements are of type
$a$.  For convenience, we often use the keyword \<\hskwd{type}\> to introduce
a new type abbreviation. For example: 
\begin{tabbing}\texfamily
{\bfseries type}~{\itshape PA}~=~{\itshape Complex}~{\itshape Double}
\end{tabbing}
introduces the new type \<PA\> as an abbreviation of the more verbose
\<Complex Double\>. A family of types that supports related operations can be
grouped in a Haskell \<\hskwd{class}\>. Individual types can then be made an
\<\hskwd{instance}\> of the class, and arbitrary code can require that a
certain type be a member of a given class.

The syntax of Haskell expressions is usually self-explanatory except perhaps
for the following points. A function can be written in at least two
ways. Both the following definitions define a function which squares its
argument:
\begin{tabbing}\texfamily
sq~n~=~n~*~n~\\
\texfamily sq'~=~\char'10~n~\char'31~n~*~n~
\end{tabbing}
A function \<f\> can be applied to every element of a list using \<map\> or
using \emph{list comprehensions}. If \<xs\> is the list \<[1,2,3,4]\>, then 
both the following:
\begin{tabbing}\texfamily
map~sq~xs~\\
\texfamily [~sq~x~|~x~$\leftarrow$~xs~]
\end{tabbing}
evaluate to \<[1,4,9,16]\>. 

Usually, a function \<f\> is applied to an argument \<a\>, by writing 
\<f a\>. If the function expects two arguments, it can either be applied to both
at once \<f (a,b)\> or one at a time \<f a b\> depending on its type. 
When convenient the function symbol can be placed between the arguments 
using back quotes \<a `f` b\>.

%%%%%%%%%%%%%%%%%%%%%%%%%%%%%%%%%%%%%%%%%%%%%%%%%%%%%%%%%%%%%%%%%%%%%%%%%%%%%%%%
\section{Proof of Monad Laws for Vectors.}
\label{sec:app2}

Proof of Proposition~\ref{prop:vec}: The definitions of \<return\> and
\<\hsbind\> satisfy the three monad laws:

\begin{itemize}
\item First monad law: \<(return x) \hsbind f = f x \>

\bhc
(return x) \hsbind f \hsalign{&= \lambda b .  sum [ return x a * f a b | a \leftarrow basis]\\
&= \lambda b . sum [ if x \hseq a then 1 else 0 * f a b | a \leftarrow basis]\\
& = \lambda b .  f x b \\
& = f x}
\ehc

\item  Second monad law: \< m \hsbind return = m\>

\bhc
m \hsbind return \hsalign{&= 
    \lambda b . sum [ m a * return a b | a \leftarrow basis]\\
& = \lambda b . sum [ m a * if a \hseq b then 1 else 0 |  a \leftarrow basis]\\
& = \lambda b . m b \\
& = m}
\ehc

\item  Third monad law: \< (m \hsbind f) \hsbind g = m \hsbind (\lambda x . f x \hsbind g)\>

\bhc
(m \hsbind f) \hsbind g \hsalign{&= (\lambda b .  sum [m a * f a b | a \leftarrow basis]) \hsbind g\\
&= \lambda c . sum[(sum [m a * f a b | a \leftarrow basis]) * g b c |\\
&\;\;\;\;\;\;\; b \leftarrow basis]\\
&= \lambda c . sum [m a * f a b  * g b c | a \leftarrow basis, b \leftarrow basis]}\\
\ \\
m \hsbind (\lambda x . f x \hsbind g)\hsalign{& = \lambda c . sum[m a * (f a \hsbind g) c | a \leftarrow basis]\\
& = \lambda c .sum[m a * (sum[f a b * g b c |  b \leftarrow basis]) |\\
&\;\;\;\;\;\;\; a \leftarrow basis] \\
& = \lambda c .sum [m a * f a b  * g b c | a \leftarrow basis, b \leftarrow basis]}
\ehc

\end{itemize}

%%%%%%%%%%%%%%%%%%%%%%%%%%%%%%%%%%%%%%%%%%%%%%%%%%%%%%%%%%%%%%%%%%%%%%%%%%%%%%%%
\section{Proof of Arrow Laws for Superoperators}
\label{sec:app3}

Proof of Proposition~\ref{prop:super}: 

\begin{itemize}
\item First arrow equation: \<arr id \ggg f = f\>. 

\bhc
arr id \ggg f \hsalign{\hsalign{ &= fun2lin (\lambda (a1,a2).(id a1, id a2)) `o` f\;\; & (by arr and \ggg)\\
&= fun2lin id `o` f & (by simplification)\\
&= return `o` f & (by fun2lin)\\
&= \lambda a . return a \hsbind f & (by `o`)\\
&= \lambda a . f a & (by monad law 1.)\\
&= f}}
\ehc

\item Second arrow equation: \<f \ggg arr id = f\>.  

\bhc
f \ggg arr id \hsalign{\hsalign{&= f `o` fun2lin (\lambda (b1,b2) . (id b1, id b2))\;\;& (by arr and \ggg)\\
&= f `o` fun2lin id  & (by simplification) \\
& = f `o` return& (by fun2lin)\\
& = \lambda a . f a \hsbind return & (by o)\\
& =\lambda a . f a & (by monad law 2.)\\
& = f }}
\ehc

\item  Third arrow equation: \<(f \ggg g) \ggg h = f \ggg (g \ggg h)\>. 

\bhc
(f \ggg g) \ggg h \hsalign{\hsalign{&=  (f `o` g) `o` h \;\;& (by \ggg)\\ 
&= \lambda b . (\lambda a . f a \hsbind g) b \hsbind h  \;\;& (by o)\\
&= \lambda b . (f b \hsbind g) \hsbind h & (by \beta)}} 
\ehc

\bhc
f \ggg (g \ggg h )\hsalign{\hsalign{&= f `o` (g `o` h) \;\;& (by \hsbind)\\
&= \lambda a . f a \hsbind (\lambda b . g b  \hsbind h)\;\;& (by o)\\
& = \lambda a  . (f a \hsbind g) \hsbind h \;\;& (by monad law 3.)}}
\ehc

\item Fourth arrow equation: \< arr (g . f) = arr f \ggg arr g\>.

\bhc
arr (g . f) \hsalign{\hsalign{& = fun2lin (\lambda (b1,b2) .((g . f) b1, (g . f) b2)) \;\;
& (by  arr)\\
&=  return .(\lambda (b1,b2) . ((g . f) b1, (g . f) b2)) &(by fun2lin)\\
&=  \lambda (b1,b2) . return ((g . f) b1, (g . f) b2) & (simplification)}}
\ehc

\bhc
arr f \ggg arr g \hsalign{& = fun2lin (\lambda (b1,b2) . (f b1, f b2)) `o` fun2lin
(\lambda (b1,b2) . (g b1, g b2))\\
&\;\;\;\;\;\; (by \hsbind and arr)\\
&= return . (\lambda (b1,b2) . (f b1, f b2)) `o` return .
(\lambda (b1,b2) . (g b1, g b2))\\
&\;\;\;\;\;\; (by fun2lin)\\
&= \lambda (b1,b2) . return (f b1, f b2) \hsbind
\lambda (b1,b2) . return (g b1, g b2))\\
&\;\;\;\;\;\; (by o)\\
&=  \lambda (b1,b2) . (\lambda (b1,b2) . return (g b1, g b2)) (f b1, f b2) \\
&\;\;\;\;\;\; (by monad law 1.)\\
&= \lambda (b1,b2) . return ((g . f) b1, (g . f) b2) \;\;(by \beta)}
\ehc

\item  Fifth arrow equation: \< first (arr f) = arr (f \times id)\>.

\bhc
first (arr f) \hsalign{&= first (fun2lin (\lambda (b1,b2) . (f b1, f b2))) \;\;\;\;
(by arr)\\
&= first (return . (\lambda (b1,b2) . (f b1, f b2))) \;\;\;\;(by fun2lin)\\
&= first (\lambda (b1,b2) . return (f b1, f b2)) \;\;\;\;(by simplification)\\
&= \lambda ((b1,d1),(b2,d2)) . \lambda ((x,y),(w,z)) . 
return (f b1, f b2) (x,w) * \\ 
& \;\;\;\;\; return (d1,d2) (y,z) \hspace{1.5cm} (by first)\\
&=  \lambda ((b1,d1),(b2,d2)) . \lambda ((x,y),(w,z)) . \\
&\;\;\;\;\;\;\hskwd{if} ((f b1,f b2), (d1, d2)) ==
((x,w),(y,z)) \hskwd{then} 1 \hskwd{else} 0 \;\;\; (by return)}
\ehc

\bhc
arr (f \times id) = \hsalign{ \hsalign{&= fun2lin (\lambda  ((b1,d1),(b2,d2)). ((f b1, d1),(f b2, d2)))
\;\;& (by arr)\\
&= return . (\lambda  ((b1,d1),(b2,d2)). ((f b1, d1),(f b2, d2))) & (by fun2lin)\\
&=  \lambda  ((b1,d1),(b2,d2)) . return ((f b1, d1),(f b2, d2)) \\
& = \lambda ((b1,d1),(b2,d2)) . \lambda ((x,y),(w,z)) . \\
&\;\;\;\;\;\;\hskwd{if} ((f b1,d1), (f b2,d2)) ==
((x,y),(w,z)) \hskwd{then} 1 \hskwd{else} 0 \;\;\;& (by return)}}
\ehc

\item Sixth arrow equation:  \< first (f \ggg g) = first f \ggg first g\>. In the following proofs
assume: \<ad1 ((b1,d1),(b2,d2)) = (b1,b2) \> and  \<ad2 ((b1,d1),(b2,d2)) = (d1,d2) \>.

\bhc
first (f `o` g) \hsalign{&= first (\lambda a . f a \hsbind g) \\
&\;\;\;\;\;\;(by `o`)\\
&= \lambda b . \lambda ((x,y),(w,z)) . (f (ad1 b) \hsbind g) (x,w) * return (ad2 b) (y,z) \\
&\;\;\;\;\;\; (by first)\\
&= \lambda b . \lambda ((x,y),(w,z)) . (\lambda c . sum [(f (ad1 b)) a * g a c | a \leftarrow basis]) (x,w) \\
&\;\;\;\;\;  * return (ad2 b) (y,z) \;\;\; (by \hsbind)\\
&= \lambda b . \lambda ((x,y),(w,z)) . sum[(f (ad1 b)) a * g a (x,w) | a \leftarrow basis] * \\
&\;\;\;\;\;  return (ad2 b) (y,z) \;\;\;(by \beta)}
\ehc 

\bhc
first f `o` first g \hsalign{&= \lambda a . first f a \hsbind \lambda b . first g b \\
&\;\;\;\;\;\; (by `o`)\\
&= \lambda a . \lambda ((x,y),(w,z)) . f (ad1 a) (x,w) * return (ad2 a) (y,z) \hsbind\\
&\;\;\;\;\; \lambda b . \lambda ((x,y),(w,z)) . g (ad1 b) (x,w) * return (ad2 b) (y,z) \\
&\;\;\;\;\; (by first)\\
&= \lambda a . \lambda ((x,y),(w,z)) . sum [ f (ad1 a) (m,o) * return (ad2 a) (n,p) * \\
&\;\;\;\;\;(\lambda ((x,y),(w,z)) . g (m,o) (x,w) * return (n,p) (y,z))((x,y),(w,z)) | \\
&\;\;\;\;\; ((m,n),(o,p))\leftarrow basis]  \;\;\;\;   (by \hsbind)\\
&= \lambda a . \lambda ((x,y),(w,z)) . sum [ f (ad1 a) (m,o) * return (ad2 a) (n,p) * \\
&\;\;\;\;\; g (m,o) (x,w) * return (n,p) (y,z)  |((m,n),(o,p))\leftarrow basis] \\
&= \lambda a . \lambda ((x,y),(w,z)) . sum [ f (ad1 a) a1 * g a1 (x,w) * \\
&\;\;\;\;\; return (ad2 a) a2 * return a2 (y,z) | a1 \leftarrow basis , a2 \leftarrow basis] \\
&\;\;\;\;\; (by simplification)\\
&= \lambda a . \lambda ((x,y),(w,z)) . sum [ f (ad1 a) a1 * g a1 (x,w) | a1 \leftarrow basis ]\\
&\;\;\;\;\; * return (ad2 a) (y,z) \;\;\; (by simplification) }
\ehc

\item Seventh arrow equation: \< first f \ggg arr (id \times g) = arr (id \times g) \ggg first f\>.

\[
lhs = first \;f \;`o` \;arr \;(id \times g)
\]

\bhc
lhs\hsalign{&= \lambda ((a1,b1),(a2,b2)) . first f ((a1,b1),(a2,b2)) \hsbind \\
&\;\;\;\;\; fun2lin (\lambda ((a,b),(c,d)) . ((a, g b),(c, g d))) \;\;
(by `o` and arr)\\
&= \lambda ((a1,b1),(a2,b2)) . first f ((a1,b1),(a2,b2)) \hsbind \\
&\;\;\;\;\; \lambda ((a,b),(c,d)) . return ((a, g b),(c, g d)) \;\;
(by fun2lin)\\
&= \lambda ((a1,b1),(a2,b2)) . \lambda ((x,y),(w,z)) . f (a1,a2) (x,w) * return (b1,b2) (y,z) 
\hsbind \\
&\;\;\;\;\;\lambda ((a,b),(c,d)) . return ((a, g b),(c, g d)) \;\;\;\;\; (by first) \\
& =\lambda ((a1,b1),(a2,b2)) . \lambda c . sum [ f (a1,a2) (m,o) * return (b1,b2) (n,p) * \\
&\;\;\;\;\; return   ((m, g n),(o, g p)) c | ((m,n),(o,p)) \leftarrow basis] \;\;\;(by \hsbind) \\
& = \lambda ((a1,b1),(a2,b2)) . \lambda ((x,y),(w,z)) . sum [ f (a1,a2) (m,o) * return (b1,b2) (n,p) * \\
&\;\;\;\;\; return   ((m, g n),(o, g p)) ((x,y),(w,z))| ((m,n),(o,p)) \leftarrow basis] \\
&\;\;\;\;\; (by simplification)\\
& = \lambda ((a1,b1),(a2,b2)) . \lambda ((x,y),(w,z)) . sum [ f (a1,a2) (m,o) * \\ 
&\;\;\;\;\; [\hskwd{if} (b1,b2) == (n,p) \hskwd{then} 1 \hskwd{else} 0] * \\ 
&\;\;\;\;\; [(\hskwd{if}  (m, g n),(o, g p)) == ((x,y),(w,z)) \hskwd{then} 1 \hskwd{else} 0] | ((m,n),(o,p)) \leftarrow basis] \\
&\;\;\;\;\; (by return)\\
&=\lambda ((a1,b1),(a2,b2)) .\lambda ((x,y),(w,z)) . \hskwd{if} (g b1, g b2) == (y,z) \\
&\;\;\;\;\; \hskwd{then}\;\;  f (a1,a2) (x,w) \hskwd{else} 0  }\\
\ehc

\[
rhs =  arr \;(id \times g)\; `o` \;first \;f
\]

\bhc
rhs \hsalign{&= \lambda ((a1,b1),(a2,b2)) . fun2lin (\lambda ((a,b),(c,d)) . ((a, g b),(c, g d))) \\
&\;\;\;\;\; ((a1,b1),(a2,b2)) \hsbind  first f \;\;\;(by `o` and arr)\\
& = \lambda ((a1,b1),(a2,b2)) . return ((a1, g b1),(a2, g b2)) \hsbind  first f\\
&\;\;\;\;\; (by fun2lin)\\
& = \lambda ((a1,b1),(a2,b2)) . first f ((a1, g b1),(a2, g b2)) \;\;\;\;(by monad law 1.)\\
&= \lambda ((a1,b1),(a2,b2)) .\lambda ((x,y),(w,z)) . f (a1,a2) (x,w) * return (g b1, g b2) (y,z) \\
&\;\;\;\;\; (by first)\\
&=\lambda ((a1,b1),(a2,b2)) .\lambda ((x,y),(w,z)) . f (a1,a2) (x,w) *\\
&\;\;\;\;\; [\hskwd{if} (g b1, g b2) == (y,z) \hskwd{then} 1 \hskwd{else} 0] \;\;\;(by return)}
\ehc

\item Eighth arrow equation: \< first f \ggg arr fst = arr fst \ggg f\>.

\[
lhs = \;first \;f \;`o` \;arr (\lambda (a,b) . a)
\]

\bhc
lhs \hsalign{&= \lambda ((a1,b1),(a2,b2)) . first f ((a1,b1),(a2,b2)) \hsbind arr \lambda (a,b) . a 
\;\;(by o)\\
&= \lambda ((a1,b1),(a2,b2)) . first f ((a1,b1),(a2,b2)) \hsbind \lambda ((a,b),(c,d)) . return (a,c) \\
&\;\;\;\;\;\; (by arr )\\
&= \lambda ((a1,b1),(a2,b2)) . \lambda ((x,y),(w,z)) . f (a1,a2) (x,w) * \\
&\;\;\;\;\;  return (b1,b2) (y,z) \hsbind \lambda ((a,b),(c,d)) . return (a,c) \;\;(by first )\\
&= \lambda ((a1,b1),(a2,b2)) . \lambda (c1,c2) . sum [ f (a1,a2) (m,o) * return (b1,b2) (n,p)*\\
&\;\;\;\;\;\; return (m,o) (c1,c2) | ((m,n),(o,p)) \leftarrow basis] \;\;\; (by \hsbind)\\
&=  \lambda ((a1,b1),(a2,b2)) . \lambda (c1,c2) .  sum [ f (a1,a2) (m,o) * \\
&\;\;\;\;\; [\hskwd{if} (b1,b2) == (n,p) \hskwd{then} 1 \hskwd{else} 0] *\\
& \;\;\;\;\; [\hskwd{if} (m,o) == (c1,c2) \hskwd{then} 1 \hskwd{else} 0] | ((m,n),(o,p)) \leftarrow basis] \;\; (by return)\\
&= \lambda ((a1,b1),(a2,b2)) . \lambda (c1,c2) . f (a1,a2) (c1,c2)\;\; (by simplification)}
\ehc

\[
rhs = arr \;fst \; `o` f
\]

\bhc
rhs \hsalign{&= \lambda ((a,b),(c,d)) . return (a,c) `o` f \;\; (by arr)\\
&=\lambda ((a1,b1),(a2,b2)) . (\lambda ((a,b),(c,d)) . return (a,c)) ((a1,b1),(a2,b2)) \hsbind f\\
&\;\;\;\;\; (by o)\\
&= \lambda ((a1,b1),(a2,b2)) . f (a1,a2) \;\;(by monad law 1.)\\
&= \lambda ((a1,b1),(a2,b2)) . \lambda (c1,c2) . f (a1,a2) (c1,c2)  }
\ehc

\item Ninth arrow equation: \< first (first f) \ggg arr assoc = arr assoc \ggg first f\>

\[
\begin{array}{l}
lhs = \lambda (((a1,b1),c1),((a2,b2), c2)) . first (first f) (((a1,b1),c1),((a2,b2), c2)) \hsbind\\
\;\;\;\;\;\;\;\; arr (\lambda ((a,b),c) . (a,(b,c)))
\end{array}
\]

\bhc
lhs \hsalign{&= \lambda (((a1,b1),c1),((a2,b2), c2)) . first (\lambda b . \lambda ((x,y),(w,z)) . 
f (ad1 b) (x,w) * \\
&\;\;\;\;\;return (ad2 b) (y,z)) (((a1,b1),c1),((a2,b2), c2)) \hsbind \\
&\;\;\;\;\;\lambda (((a1,b1),c1),((a2,b2), c2)) . return ((a1,(b1,c1)),(a2,(b2,c2))) \\
&\;\;\;\;\; (by first)\\
&= \lambda (((a1,b1),c1),((a2,b2), c2)) .  \lambda ((m1,n1),p1) ((m2,n2), p2) . \\
&\;\;\;\;\; (\lambda b . \lambda ((x,y),(w,z)) . f (ad1 b) (x,w) * return (ad2 b) (y,z)) ((a1,b1),(a2,b2)) \\
&\;\;\;\;\; ((m1,n1),(m2,n2)) * return (c1,c2) (p1,p2) \hsbind \lambda (((a1,b1),c1),((a2,b2), c2)) . \\
&\;\;\;\;\;return ((a1,(b1,c1)),(a2,(b2,c2))) \;\;\;\;\; (by first)\\
&= \lambda (((a1,b1),c1),((a2,b2), c2)) .  \lambda ((m1,n1),p1) ((m2,n2), p2) . \\
&\;\;\;\;\;f (a1,a2) (m1,m2) * return (b1,b2) (n1,n2) *  return (c1,c2) (p1,p2) \hsbind \\
&\;\;\;\;\;\lambda (((a1,b1),c1),((a2,b2), c2)) . return ((a1,(b1,c1)),(a2,(b2,c2)))\\
&\;\;\;\;\; (by \beta)\\
&=\lambda (((a1,b1),c1),((a2,b2), c2)) . \lambda ((x1,(y1,z1)),(x2,(y2,z2))) . \\
&\;\;\;\;\;  sum [ f (a1,a2) (m1,m2) * return (b1,b2) (n1,n2) * return (c1,c2) (p1,p2) * \\
&\;\;\;\;\; return  ((m1,n1),p1) ((m2,n2), p2) ((x1,(y1,z1)),(x2,(y2,z2))) |\\
&\;\;\;\;\;((m1,n1),p1) ((m2,n2), p2) \leftarrow basis] \\
&\;\;\;\;\; (by \hsbind)\\
&=\lambda (((a1,b1),c1),((a2,b2), c2)) . \lambda ((x1,(y1,z1)),(x2,(y2,z2))) . \\
&\;\;\;\;\;  sum [ f (a1,a2) (m1,m2) *
[\hskwd{if} (b1,b2) == (n1,n2) \hskwd{then} 1 \hskwd{else} 0] * \\
&\;\;\;\;\; [\hskwd{if} (c1,c2) == (p1,p2)\hskwd{then} 1 \hskwd{else} 0]  * \\
&\;\;\;\;\; [\hskwd{if} ((m1,n1),p1) ((m2,n2), p2) == ((x1,(y1,z1)),(x2,(y2,z2))) \hskwd{then} 1 \hskwd{else} 0]|\\
&\;\;\;\;\;((m1,n1),p1) ((m2,n2), p2) \leftarrow basis] \\
&\;\;\;\;\; (by return)\\
&=  \lambda (((a1,b1),c1),((a2,b2), c2)) . \lambda ((x1,(y1,z1)),(x2,(y2,z2))) . f (a1,a2) (x1,x2) * \\
&\;\;\;\;\; return ((b1,c1),(b2,c2)) ((y1,z1),(y2,z2))}
\ehc

\[
rhs = \lambda (((a1,b1),c1),((a2,b2), c2))\; . \;return ((a1,(b1,c1)),(a2,(b2,c2))) \;`o` \;first \;f
\]

\bhc
rhs \hsalign{&= \lambda (((a1,b1),c1),((a2,b2), c2)) . return ((a1,(b1,c1)),(a2,(b2,c2))) \hsbind first f\\
&\;\;\;\;\;\;(by o)\\
&=  \lambda (((a1,b1),c1),((a2,b2), c2)) . first f ((a1,(b1,c1)),(a2,(b2,c2)))\\
&\;\;\;\;\;\; (by monad law 1.)\\ 
&=  \lambda (((a1,b1),c1),((a2,b2), c2)) . \lambda ((x1,(y1,z1)),(x2,(y2,z2))) .\\
&\;\;\;\;\; f (a1,a2) (x1,x2) * 
return ((b1,c1),(b2,c2)) ((y1,z1),(y2,z2)) \;\;\;(by first) }
\ehc

\end{itemize}

%%%%%%%%%%%%%%%%%%%%%%%%%%%%%%%%%%%%%%%%%%%%%%%%%%%%%%%%%%%%%%%%%%%%%%%%%%%%%%%%

\bibliographystyle{plain}
\bibliography{arXiv}

\end{document}